\documentclass[preprint]{aastex}

\shorttitle{NOVAE IN M94}
\shortauthors{G\"{U}TH, SHAFTER, \& MISSELT}

\begin{document}
\title{The Nova Rate in M94 (NGC~4736)}


\author{T. G\"{u}th and A. W. Shafter}
\affil{Astronomy Department, San Diego State University, San Diego, CA 92182}
\email{tgueth@sciences.sdsu.edu, aws@nova.sdsu.edu}

\and

\author{K. A. Misselt}
\affil{Steward Observatory, University of Arizona, 933 North Cherry Avenue,
  Tucson, AZ 85721}
\email{misselt@as.arizona.edu}

\begin{abstract}
A multi-epoch H$\alpha$ survey of the early-type spiral galaxy M94 (NGC
4736) has been completed as part of a program to establish the galaxy's
nova rate. A
total of four nova candidates were discovered in seven epochs of observation
during the period from 2005 to 2007. After making corrections for temporal
coverage and spatial completeness, a global nova rate of
5.0$^{+1.8}_{-1.4}$ yr$^{-1}$ was determined. This rate
corresponds to a specific-luminosity nova
rate of 1.4 $\pm$ 0.5 novae per year per 10$^{10} L_{\odot,K}$ when the $K$
luminosity is determined from the $B-K$ color, or 1.5 $\pm$ 0.4 novae per year per
10$^{10} L_{\odot,K}$ when the $K$ luminosity is derived from the Two Micron
All Sky Survey. These values are slightly lower than that of other galaxies
with measured nova rates, which typically lie in the range of $2-3$ novae per
year per 10$^{10} L_{\odot}$ in the $K$ band. 
\end{abstract}


\section{Introduction}

Novae, a subclass of cataclysmic variables, occur in semidetached binary
systems where a white dwarf primary star accretes
material from its Roche lobe-filling late-type companion
\citep[e.g., see][]{war95}.
When the mass accretion rate is sufficiently low, the
accreted material builds up on the surface of the white dwarf under degenerate
conditions. A thermonuclear runaway (TNR) eventually ensues when the temperature
and density at the base of the accreted material become sufficiently
high. The TNR causes the accreted material to be blown off, resulting in an
outburst of 10 -- 20 mag. The corresponding peak luminosity can be
as high as $M_{V} \sim -10$, and visible in galaxies as distant as the
Virgo cluster. The properties of nova eruptions (peak luminosity and fade rate)
are expected to be strongly dependent on properties such as the mass accretion
rate, the luminosity and the mass of the WD, which are expected to change with
the underlying stellar populations \citep[e.g.][]{sha80, pri82, liv92}.


\indent In a seminal paper, \citet{yun97} presented the results of stellar
population synthesis models that predicted the infrared (e.g., $K$ band) 
luminosity-specific nova rate (LSNR)
of a galaxy should have a strong dependence on the history of its star
formation rate (SFR). This dependence is due to an expected increase in
the average mass of the white dwarfs in nova binaries
with decreasing time elapsed since the zero-age main-sequence
system formed \citep{tut95}. Since higher mass white dwarfs exhibit
more frequent, and more luminous, nova eruptions
\citep{rit91, liv92, kol95},
elliptical galaxies, which
form most of their stars in an early burst of star formation, were not
expected to be prolific nova producers. Alternatively, galaxies that have
formed a significant fraction of their stars relatively recently, like
late-type spiral galaxies, were predicted to have higher observed
LSNRs. In addition to Hubble type,
the LSNRs of galaxies were also expected to be dependent
on galaxy mass. Over their lifetimes, dwarf, late-type
spiral galaxies are believed to have an almost constant SFR,
while their giant, late-type counterparts are thought to have star
formation rates that are exponentially decaying \citep{gav96}.
Thus, massive spiral
galaxies are predicted to have less active SFRs due to low
LSNRs when compared with the dwarf spiral and irregular systems.

\indent Over the past two decades, nova surveys have been conducted
in galaxies spanning a wide
range of Hubble types. 
In an early study based on a limited sample of galaxies, \citet{cia90a}
concluded that the $K$-band LSNR was basically
independent of Hubble type. Shortly thereafter, \citet{del94} re-analyzed
many of these same galaxies and
reached a somewhat different conclusion; namely, that the LSNRs are
systematically higher in the late-type, low-mass
spiral galaxies, such as M33, the Large Magellanic Cloud (LMC), and
the Small Magellanic Cloud (SMC), consistent with the
predictions of the synthesis models subsequently published
by \citet{yun97}.
A principal limitation of these early studies was the lack of data for massive,
late-type spiral galaxies for which nova rates had yet to be determined.
With this in mind,
\citet{sha00} surveyed two massive, late-type
spirals, M51 and M101, and, for comparison, the giant Virgo elliptical
galaxy, M87, finding a slightly higher
LSNR for M87 than for either of the massive spirals M51 or M101
(the difference, however, was not statistically significant).
In more recent years several additional nova
surveys have been conducted \citep{sha02, fer03, wil04, coe08}.
Like \citet{sha00}, most of these studies failed to find
any convincing variation
of the LSNR with Hubble type. However, large uncertainties
in the LSNRs for many galaxies remain, with, for example,
published nova rates for
the giant Virgo elliptical M87 varying by as much as a factor of 3
\citep{sha00, sha02}.


\indent To make further progress in understanding what effect, if any, the
underlying stellar population might have on the LSNR it is important to
obtain nova rates for additional galaxies spanning a broad range of
Hubble types. With this motivation
in mind, we have conducted a survey for novae in the early-type Sab galaxy M94
(NGC 4736), which is well known for its inner ring of
ongoing starburst activity \citep{wal01}. Here we
report the results of our survey.

\section{Observations}

Data for M94 were obtained over seven epochs spanning 2005 May to 2007 June using
the Steward Observatory 2.3-m Bok Telescope.  All observations were made with
the 90Prime camera \citep{will04}, consisting of an array of four 4K
$\times$ 4K CCDs mounted at prime focus, giving a full field of view of $\sim$1
deg$^{2}$.  For our observations,
the $\sim$ $30'\times30'$ coverage provided
by a single chip was sufficient to cover essentially the full disk of M94.

\indent As emphasized by \citet{cia90b} conducting nova surveys
in H$\alpha$ has several advantages over observations
in the broadband continuum.  Shortly after eruption, novae
develop strong and broad ($\gtrsim$ 1000 km s$^{-1}$) H$\alpha$ emission lines
that fade slowly, normally requiring several months to decline by more than 2
mag. Thus, novae can be detected in H$\alpha$ with less frequent temporal
sampling. In addition, observations in H$\alpha$ are also less affected by
extinction than those in $B$ light, which provides an added advantage
for detection of novae in spiral galaxies. Finally, the narrow-band
H$\alpha$ observations
facilitate the detection of novae against a bright
background, particularly near the nuclei of galaxies.
For these reasons, following our earlier work \citep[e.g.][]{sha00,
coe08}, observations of M94
were made using a narrow-band H$\alpha$ filter having a central
bandpass of 6580\ \AA \ and a FWHM of $\sim$80\ \AA. 

\indent Based on the distance modulus to M94, which we take to be
$\mu_{0} = 28.21 \pm 0.07$ (Herrmann et al. 2008),
we estimated that a total exposure time
of $\sim$2~hr would allow us to reach a limiting absolute magnitude of
$M_{\rm H\alpha} \sim -6.7$ ($m_{\rm H\alpha} \sim 21.5$) necessary
to detect a significant fraction of the
novae. Our target exposure time of 2~hr was divided
into a series of individual 600--900~s exposures that were later
median stacked in order to suppress cosmic ray artifacts and
to avoid saturation of the brightest stars in the image. In addition,
a set of sky flats, bias frames, and dark frames were obtained
for each epoch of observation.
The individual images were first de-biased and flat-fielded
before a world coordinate system (WCS) was assigned to each image with the
IRAF\footnotemark[1] routine \texttt{msctpeak} using coordinates from the US
Naval Observatory A2 catalog.
The images of a given epoch were then aligned using several stars common to
each image.  In four of the seven epochs,
the galaxy was centered on a single chip (chip 1)
in all exposures, allowing the images to be aligned with the IRAF routine
\texttt{imalign}.  In the remaining three epochs, exposures of
M94 were alternated between three CCDs (chip 3 was excluded due to a large
area of bad pixels). As a result, use of
the IRAF routine \texttt{sregister}, which accounts for the slight
rotation from one chip to another, was required to align these images.
Individual images from a given observing run (whether a single night or two
consecutive nights such as 2005 May 1 and 2, and 2005 May 30 and 31)
were then combined into a single master image for that epoch.
The resulting master images for each of seven epochs ended up representing between
1.67 and 3 hr of coverage. A summary of the observations can be
found in Table~1.

\footnotetext[1]{IRAF (Image Reduction and Analysis Facility) is distributed
  by the National Optical Astronomy Observatory, which is operated by
  Association of Universities for Research in Astronomy, Inc., under
  cooperative agreement with the National Science Foundation} 

\subsection{Nova Detection}


In order to search for novae most efficiently, the co-added images for all
epochs separated by more than one year were spatially aligned and
point-spread function (PSF) matched
using the ISIS package \citep{ala98}. The resulting images
were differenced in order to search for variable objects.
To then qualify as a nova, an object had to appear in at least
one epoch but not be present in other images obtained more than 6 months
before or after the epoch of detection.

To confirm our nova candidates, and to search for any additional candidates
that might have been missed, the images were also blinked by eye. Owing
to the bright background galaxy light this procedure was ineffective
within $\sim1.5\arcmin$ from the nucleus. For this inner region,
we employed the IRAF \texttt{median} routine
with a 7 $\times$ 7 pixel box to generate smoothed background images
that were then subtracted from their original images.
These median subtracted images were then also blinked by eye.
All four candidates were confirmed by the visual blinking process,
but no further candidates were identified.

\subsection{Nova Photometry}

Photometry was performed using two different techniques. First, using the
IRAF routine \texttt{psf} from the DAOPHOT package, raw magnitudes of the nova
candidates were obtained.  To confirm these measurements, the second IRAF
routine used was \texttt{phot}, which is a simple aperture photometry method.
We first had to extract 100 $\times$ 100 sub-images centered on each nova and
standard star due to the bright galactic background.  Next, we excluded a 5
pixel radius aperture around the object and used the IRAF routine
\texttt{imsurfit} to fit a two-dimensional surface to the background, which
was then subtracted from the image. This left a flattened image on which
aperture photometry was performed. Using apertures with diameter on the order
of the seeing FWHM, we obtained the instrumental magnitudes for both the nova
candidates and for a secondary standard star we defined, which is located at
RA = 12hr 50m 48s.07, DECL = 41$^{\circ}$10$\arcmin$06$\arcsec$.44 (J2000).
The secondary star was later calibrated ($m_{\rm H\alpha}$ = 15.71) by comparison
with the spectrophotometric standard star HZ~44 \citep{oke74} when both stars
were observed under photometric conditions on the night of 2008 June 3 UT.
This enabled us to convert the instrumental magnitudes of the novae to
H$\alpha$ magnitudes. In general, the magnitudes acquired from both the
\texttt{psf} and \texttt{phot} methods were in agreement to within 0.1
mag. Hence, we used the mean value from the two readings unless they differed
by more than 0.1 mag, in which case the \texttt{psf} value was adopted. In
this study, we made the assumption that the H$\alpha$ filter has a filling
fraction of 100$\%$ as it has been done by \citet{sha01}, \citet{wil04}, and
most recently by \citet{coe08}.  The H$\alpha$ emission line width changes
from nova to nova, hence the bandpass can be underfilled, introducing a small
error in the calibration. Nevertheless, making this assumption allows direct
comparison with previous studies.

\indent Table~2 lists the dates, positions, and H$\alpha$ magnitudes of the
four nova candidates discovered in this survey.  M94N~2006-04a and M94N
2006-04b were detected in multiple epochs, but we were unable to constrain
the times of maximum light of these novae. Problems with the detector rendered
our measurements from 2006 May uncertain.  Nevertheless, our observations
indicated that both of these nova candidates faded quite slowly.
M94N~2006-04a faded by only about 0.5 mag in 60 days,
raising the possibility that this transient source may not be a nova,
but rather a luminous, long-period variable star.
M94N~2006-04b is similar, but is more likely to be a slow
nova considering that it declined faster and
was almost a magnitude brighter at discovery, corresponding to
an absolute magnitude of $M_{\rm H\alpha}\simeq-8.2$ at the distance of M94.

\section{The Nova Rate in M94}

\begin{figure}
\centering
\includegraphics{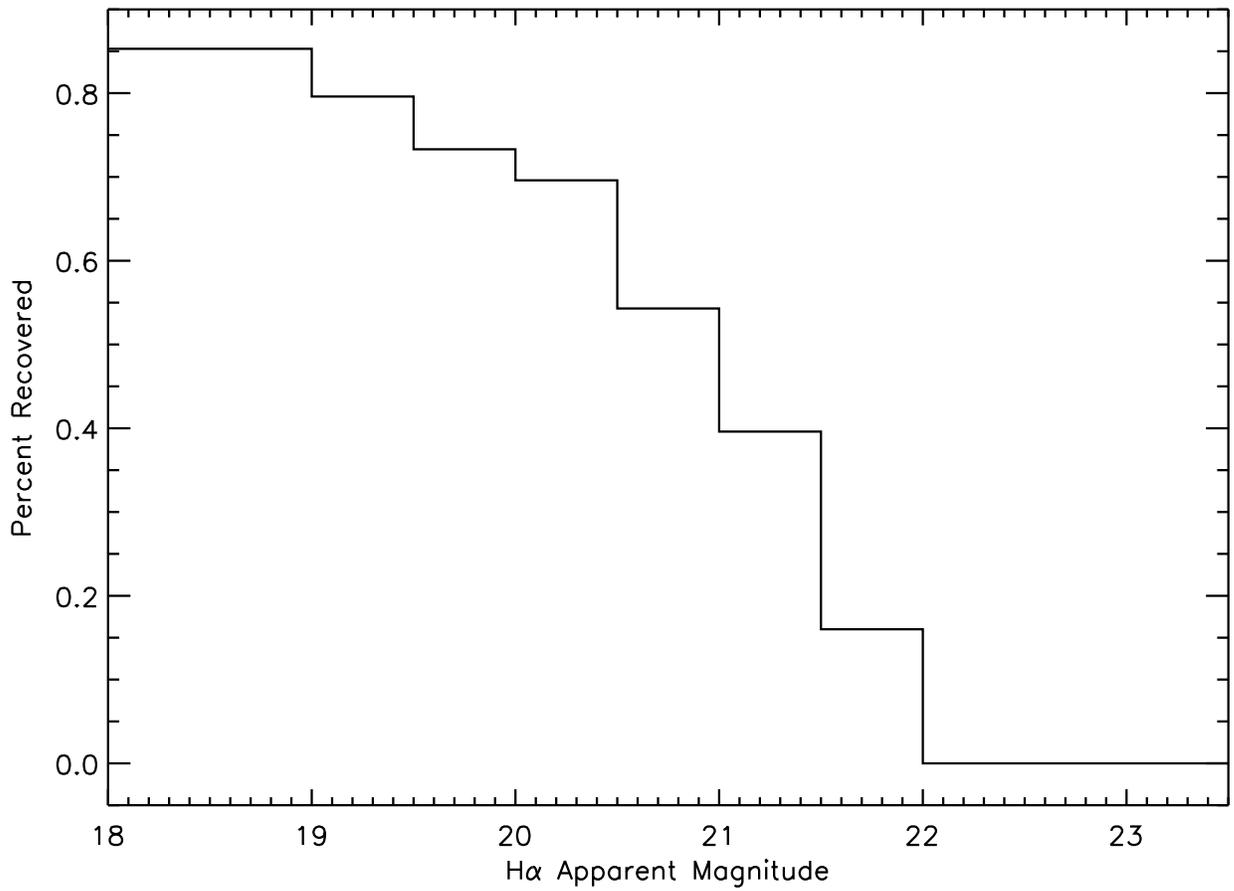}
\caption{Fraction of artificial novae recovered during artificial star tests
  as a function of magnitude, $C(m)$.  The limiting magnitude of $m_{\rm H
  \alpha}$ = 21.0, which is used in the mean lifetime nova rate calculation,
  is determined by the steep drop-off in completeness at fainter magnitudes.}
\label{fig:LimitingMag}
\end{figure}

To estimate the nova rate in M94, the completeness of the survey as a function
of magnitude must be determined.  We determined the completeness of our survey
by performing artificial star tests using the IRAF routine \texttt{addstar} as
in \citet{wil04} and \citet{coe08}. For 12 equally spaced magnitude bins,
artificial novae with magnitudes ranging between $m_{\rm H\alpha}$ = 18.0 and
$m_{\rm H\alpha}$ = 23.0 were created. Following the spatial distribution of the
$I$-band light (from the data of \citet{moe95}), 100 artificial novae were
randomly distributed throughout the image.  The search for these artificial
``novae''
followed the same methods as the ones employed to determine the actual novae.
This procedure, which was repeated three times and averaged, allowed us to
construct a completeness function, $C(m)$, of the number of novae recovered in
each magnitude bin, shown in Figure \ref{fig:LimitingMag}.  The recovered novae
fraction starts to decline steeply at magnitude of
$m_{\rm H\alpha}\simeq21.0$. Following \citet{sha00}, \citet{wil04}, and
\citet{coe08}, we obtained the nova rate in M94 by employing two different
procedures: a Monte Carlo simulation and a mean nova lifetime method.

\subsection{The Monte Carlo Procedure}

We estimated the nova rate in M94 by employing a Monte Carlo simulation where
the actual number of novae observed in this survey ($n_{obs}$ = 4) was
compared with an estimate of the number of novae we could expect to see
given an intrinsic nova rate, $R$.  For a wide range of possible values of
$R$, we computed a set of model H$\alpha$ light curves based on randomly
selected peak magnitudes and decay rates from a sample of detected novae in
earlier studies of M31 \citep{sha01} and M81 \citep{nei04}. Based on the
dates of our survey (Table~1) and our adopted distance to M94,
we then computed an observed nova luminosity
function, $n(m,R)$, for this range of possible nova rates.  The resulting
luminosity function was then convolved with the completeness function, $C(m)$,
to determine the expected number of novae in M94: 

\begin{equation}
N_{obs}(R) = \int{C(m)n(m,R)dm}
\label{eqn:NovaNum}
\end{equation}

\begin{figure}
\centering
\includegraphics{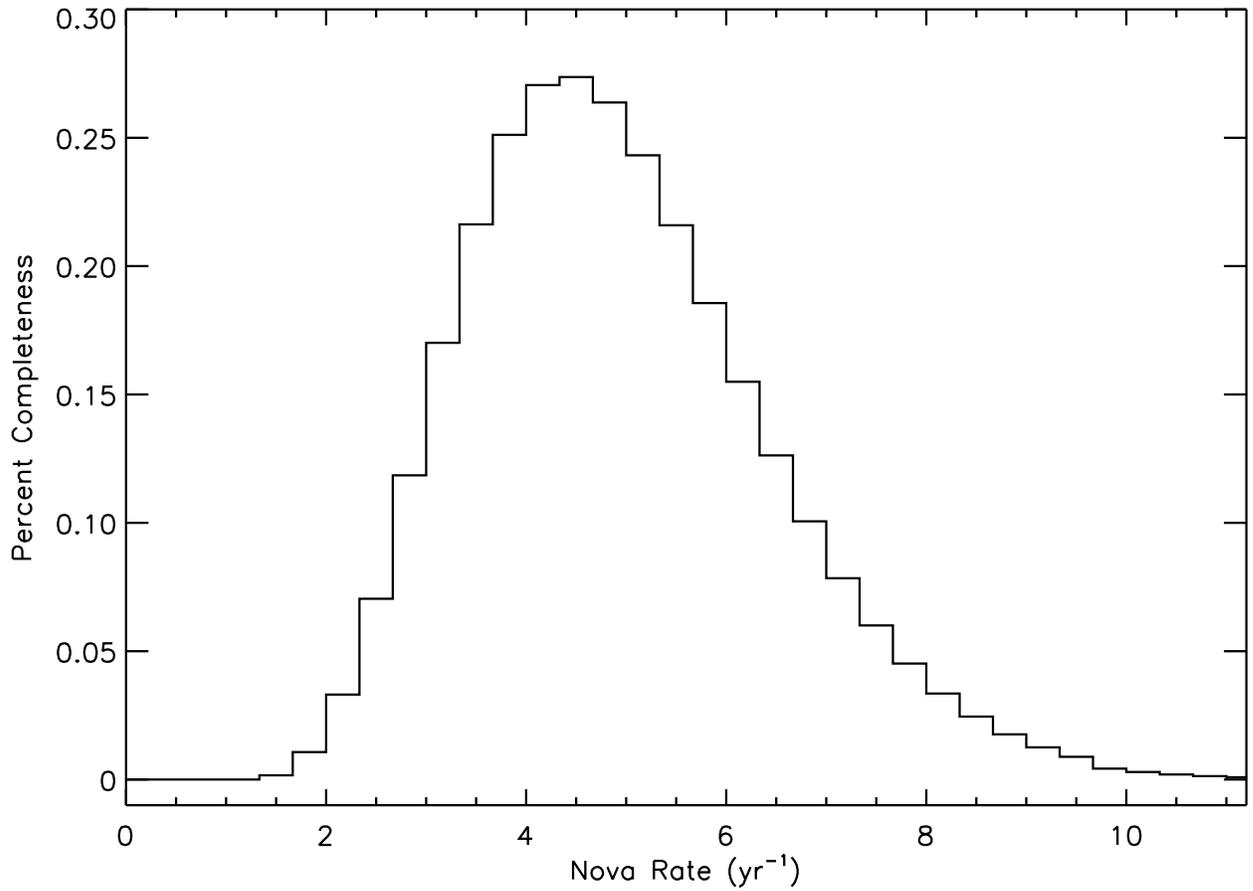}
\caption{Results from the Monte Carlo simulation for the four nova candidates. The
  peak in the normalized probability distribution ($R=4.7$~yr$^{-1}$)
  represents the most probable
  nova rate in M94 within the surveyed region of the galaxy.}
\label{fig:MonteCarlo}
\end{figure}

We repeated the Monte Carlo simulation 10$^{6}$ times and recorded the number
of matches between $N_{obs}(R)$ and $n_{obs}$ = 4, the actual number of novae
identified in this survey.  We then normalized the number of matches as a
function of $R$, and obtained the probability distribution function shown in
Figure \ref{fig:MonteCarlo}.  The most probable nova rate, represented by the
peak in the figure, is 4.7$^{+1.6}_{-1.2}$ yr$^{-1}$, where the error
range encompasses 50\% of the integrated probability distribution.
As we pointed out
earlier, M94N~2006-04a has features that makes it difficult to confirm it as a
nova candidate with complete confidence.  If we exclude M94N~2006-04a, the
nova rate drops as expected to 3.4$^{+1.6}_{-0.9}$ yr$^{-1}$. 

\subsection{The Mean Nova Lifetime Procedure}

We also
determined an estimate of the nova rate using the mean nova lifetime method
(Ciardullo et al. 1990b), which follows the original analysis
by \cite{zwi42} to determine
extragalactic supernova rates. Although this method is less sophisticated than
the Monte Carlo procedure, it can be used as a check on our earlier results.
We begin by expressing the nova rate, $R$, in the
studied region as
 
\begin{equation}
R = \frac{N(M < M_{c})}{T(M < M_{c})}
\label{eqn:NovaRate}
\end{equation}

where $N(M < M_{c})$ is the number of novae observed brighter than the
limiting absolute magnitude $M_{c}$ of the survey, and $T(M < M_{c})$ is
the ``effective survey time.''  For multi-epoch surveys such
as this survey, the effective survey time both depends on the mean nova
lifetime, $\tau_{c}$, which is defined as the length of time a common nova
remains brighter than $M_{c}$, and the frequency of sampling.  Therefore:

\begin{equation}
T(M < M_{c}) = \tau_{c} + \sum_{i=2}^{n}{\rm min}(t_{i} - t_{i - 1},\tau_{c})
\label{eqn:EffTime}
\end{equation}

where $t_{i}$ is the time of the $i$th observation.  Using observations from
the bulge of M31, \citet{sha00} provide a simple calibrated relationship
between $\tau_{c}$ and $M_{c}$, which we implement here:

\begin{equation}
\log\tau_{c}(\rm days) \simeq (6.1 \pm 0.4) + (0.56 \pm 0.05)M_{c}
\label{eqn:TauTime}
\end{equation}

Adopting a limiting magnitude of $m_{\rm H\alpha}$ = 21.0,
$\mu_{0} = 28.21 \pm 0.07$, and assuming
a foreground extinction of $\sim$0.08 mag \citep{sch98},
we estimate that $M_{c} = -7.29 \pm 0.21$. The mean nova
lifetime and the effective survey time were calculated using the survey times
provided in Table~1, and Equations (\ref{eqn:EffTime}) and
(\ref{eqn:TauTime}), which yields $\tau_{c} = 100.4 \pm 4.7$ days and $T(M <
M_{c}) = 480 \pm 14$ days, respectively. Equation (\ref{eqn:NovaRate})
provides a nova rate of 3.0 $\pm$ 1.5 yr$^{-1}$. Given that we are only
$\sim$55\% complete down to our limiting magnitude of $m_{\rm H\alpha}$ = 21.0
(see Figure \ref{fig:LimitingMag}), we estimate the nova rate in the surveyed
region to be 5.4 $\pm$ 2.7 nova per year.  This nova rate agrees, within the
margins of error, with the Monte Carlo nova rate value. 

\indent As with the Monte Carlo method, the mean nova lifetime rate was also
determined excluding the uncertain nova M94N~2006-04a. Both
the effective survey time and the mean nova lifetime are unaffected by the
change in the number of nova candidates. In this case the nova rate drops to 2.3
$\pm$ 1.3 yr$^{-1}$, or 4.1 $\pm$ 2.4 yr$^{-1}$ after correcting for the
completeness at $m_{\rm H\alpha}$ = 21.0.

\subsection{The Global Nova Rate}

\begin{figure}
\centering
\includegraphics{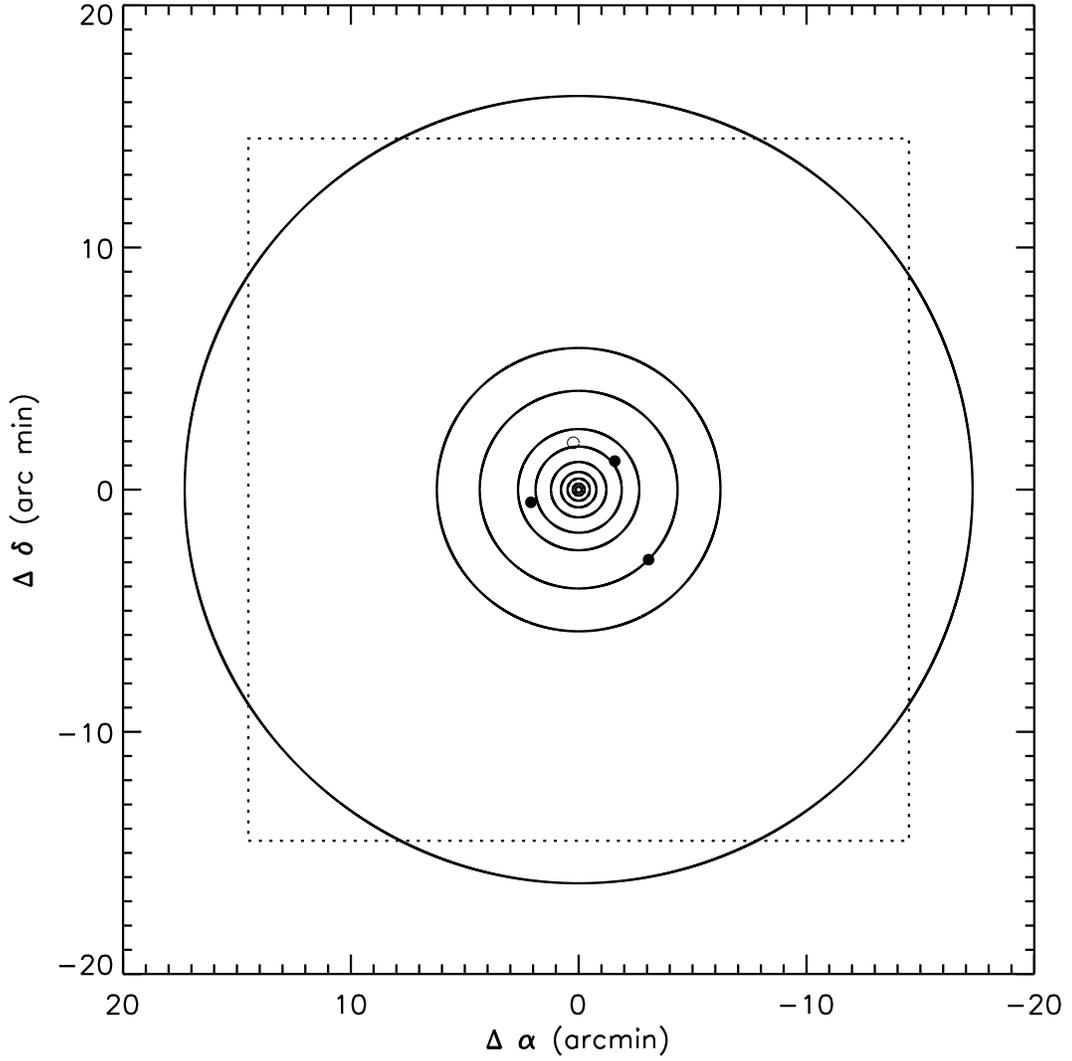}
\caption{Spatial distribution of the four nova candidates
  found in M94. All four objects
  were found to be above the limiting magnitude of $m_{\rm H \alpha}$ = 21.0.
  The open circle represents a questionable source that could possibly
  be a long-period variable star masquerading as a nova.
  Also shown are the mean $I$-band isophotes
  obtained from \citet{moe95} and the sampled region of this survey, defined
  by the dashed box centered on the nucleus of the galaxy.}
\label{fig:SpatialDistr}
\end{figure}

The large field of view of the 90Prime camera allows us to cover M94 almost
completely, which requires only a small extrapolation to estimate
the nova rate of the entire
galaxy.  Using the $I$-band photometry data of \citet{moe95} for M94, we
estimate that $\sim$95\% of the total infrared luminosity of the galaxy is
included in the effective survey area, with an estimated $\sim$5\% lost due
to the placement of the galaxy on the CCD during some epochs of observation.
Both the Monte Carlo and mean nova lifetime rates calculated above
represent the rates in the surveyed region of M94 only, which need to be
corrected for any fraction of the galaxy that falls outside the survey's
coverage.  Assuming the 95\% coverage, we obtain 5.0$^{+1.8}_{-1.4}$ yr$^{-1}$
and 5.7 $\pm$ 2.9 yr$^{-1}$ for the Monte Carlo global nova rate and mean nova
lifetime global nova rate, respectively. If M94N~2006-04a is
omitted, the estimate of the global nova rate based on the Monte Carlo and the
mean nova lifetime drop to 3.8$^{+1.9}_{-1.1}$ yr$^{-1}$ and 4.3 $\pm$ 2.5
yr$^{-1}$, respectively. These estimates have not been corrected for
any extinction internal to M94, which is highly
dependent on spatial position.
In Figure \ref{fig:SpatialDistr}, we have
plotted the spatial distribution of all four nova candidates discovered in
this survey over the $I$-band isophotes of M94 \citep{moe95}.  Each isophote
represents a 10\% change in the total light of M94. The three strong nova
candidates are represented by filled circles with the slowly fading
nova candidate M94N~2006-04a
represented by an open circle. All four nova candidates lie outside the
inner starburst ring in M94, which has a diameter of approximately
$70''$. The dotted square, which represents the size of
the surveyed region, is a 29$\arcmin \times$ 29$\arcmin$ region centered on
the nucleus of the galaxy.

\subsection{Luminosity-specific Nova Rate}

To compare nova rates across different stellar populations, nova rates are
typically normalized by the integrated infrared $K$-band luminosity of the
galaxy, yielding a $K$-band luminosity-specific nova rate (LSNR), $\nu_{K}$,
which is usually parameterized as the number of novae per year per 10$^{10}$
$L_{\odot,K}$.

\indent There are two different ways to estimate the integrated
$K$-band luminosity. The
value can be obtained directly from the Two Micron All Sky Survey's (2MASS)
Large Galaxy Atlas \citep{jar03} or it can be estimated indirectly
using the $B$ integrated magnitude and $(B-K)$ color.
Although the 2MASS data provides direct $K$ magnitude measurements
of galaxies with determined nova rates, there are systematic differences
between these values and those from the $(B-K)$ color \citep{fer03, wil04,
  coe08}.  Those discrepancies arise because the sky background levels near
large galaxies are quite hard to measure accurately as pointed out by
\citet{wil04}.  As a result, we consider the LSNR obtained from the galaxy
colors to be more dependable, as concluded by \citet{wil04} and \citet{coe08}.

\indent We adopt an integrated $B$ magnitude for
M94 of 8.99 $\pm$ 0.13 \citep{dev91}. This value, coupled with a $(B-K)$
color of 3.72 $\pm$ 0.10 \citep{joh66, dev91} yields an integrated
$K$-band magnitude of $5.18\pm0.016$. At the distance of M94,
our nova rate of 5.0$^{+1.8}_{-1.4}$ yr$^{-1}$ yields
$\nu_{K,\rm color} = 1.4 \pm 0.5 \times 10^{-10}L_{\odot,K}^{-1}$ yr$^{-1}$.
If we were to adopt the revised $K$
magnitude value of 5.11 $\pm$ 0.02 from 2MASS, we find a slightly
higher value of
$\nu_{K,\rm 2MASS} = 1.5 \pm 0.5 \times 10^{-10}L_{\odot,K}^{-1}$ yr$^{-1}$.
Finally, if we exclude M94N~2006-04a, we obtain
$\nu_{K,\rm color} = 1.0 \pm 0.4 \times 10^{-10}L_{\odot,K}^{-1}$ yr$^{-1}$ and
$\nu_{K,\rm 2MASS} = 1.1 \pm 0.4 \times 10^{-10}L_{\odot,K}^{-1}$ yr$^{-1}$ for the
galaxy colors and 2MASS estimates, respectively.  

\section{Discussion}

\begin{figure}
\centering
\includegraphics{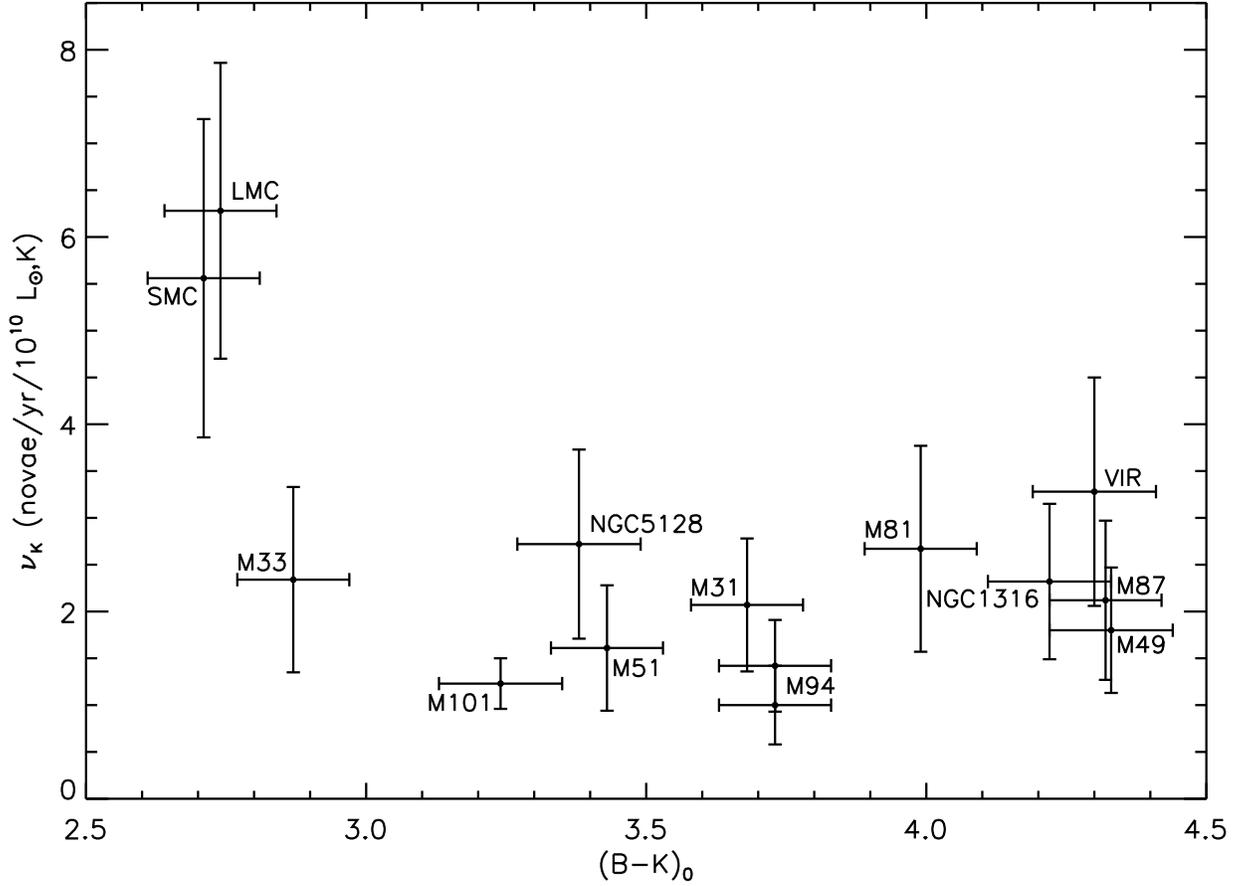}
\caption{LSNRs of surveyed galaxies are plotted as a function of the $(B-K)$
  color of the galaxy. Most galaxies have LSNRs values that are close to
  a typical value of
  $\nu_{K}$ $\sim$ (2 $\pm$ 1)$\times 10^{-10}L_{\odot,K}^{-1}$ yr$^{-1}$. We
  have plotted two points for M94 representing the LSNR based on the detection
  of either three or four novae. Note
  that the two low-mass, late-type irregular Magellanic Clouds have LSNRs that
  are $\sim$2--3 times greater.}
\label{fig:LSNR}
\end{figure}

In an attempt to broaden the study of galaxies with different Hubble types, we
have undertaken the first systematic survey for novae
in the spiral galaxy M94.
Figure \ref{fig:LSNR} shows the LSNRs for M94 and other galaxies with
measured nova rates plotted
versus the $B-K$ color of the host galaxy.
Considering
that only three of the four nova candidates in M94 may in fact be bona fide novae,
we have plotted LSNRs appropriate for both possibilities.
The data for all
galaxies, except M94, are obtained from Tables~5 and~6 of \citet{wil04}, and
include the most recent LSNR value of M101 from \citet{coe08}.


\indent Our best estimate of the
LSNR for M94 ($[1.4\pm0.5]\times$ 10$^{-10} L_{\odot,K}^{-1}$ yr$^{-1}$)
is in general agreement with the previous findings of
Ciardullo et al. (1990a) and more recently with those of
Shafter et al. (2000), Williams \& Shafter (2004), and Coelho et al. (2008),
who showed that the LSNR is remarkably insensitive to the Hubble type
of the parent galaxy. Notable exceptions are apparently the Magellanic
Clouds. The slightly higher rates for the LMC and the SMC can be broadly
understood within the context of the population synthesis models
of Yungelson et al. (1997) who argued that among late-type galaxies,
the low-mass systems with their relatively constant star formation
histories, should have generally higher LSNRs.

M94 is noted for a ring of active star formation with a diameter of
$\sim70''$ surrounding the galaxy's
bright nucleus. Given the models of Yungelson et al. (1997), which
predict a higher LSNR in galaxies with a recent history of star formation,
it is perhaps surprising that the LSNR for M94 is not higher than those
of galaxies with similar Hubble types, but with less active star formation.
In this regard, it is noteworthy
that the four nova candidates discovered in our
survey are located well outside the star-forming ring, and
appear to have no association with it. Perhaps
the relatively low LSNR for M94 and the lack of novae associated
with the ring might be due to the difficulty in
detecting nova candidates in the inner star-forming
regions of the galaxy. Extinction internal to M94 may be significant,
especially in the star-forming regions,
and our inability to correct for it could lead to an underestimation
of the nova rate.
Further, we note that
the adopted distance to M94 affects the derived LSNR.  Although we use the most
recent value by \citet{her08}, the distance to M94 is not well established. If
we have underestimated the distance to M94, our derived nova rate will
underestimate the galaxy's true nova rate.

\indent It is also possible that our limited temporal coverage might result in
an underestimation of the actual nova rate of M94.
Both the Monte Carlo
and mean nova lifetime approaches attempt to correct for survey frequency,
but both methods rely on an accurate
knowledge of the nova light curve properties.
A concern is whether the H$\alpha$ light curve data from
M31 and M81 can be used in the Monte Carlo simulations for
galaxies like M94 with differing Hubble types and SFRs. 
An H$\alpha$ survey of M81 with nearly continuous
temporal coverage over a $\sim$5 month long period conducted by \citet{nei04}
showed that the LSNR of M81 is more than twice the value of that reported
earlier by \citet{mos93}. The authors argue that multi-epoch surveys have
infrequent temporal coverage, which will systematically underestimate nova
rates. However, \citet{nei04} used a lower $K$-band luminosity for M81 that
was adopted from the 2MASS data \citep{jar03}, which will increase their
LSNR. If we adjust for this difference, the LSNR computed by
\citet{nei04} is only about $\sim$40\% greater than the preliminary value found
by \citet{mos93}.

If the slightly low LSNR for M94 is confirmed, and is not a result of
the uncertainties outlined above, it would be tempting to associate the
modest LSNR to the small bulge of M94 \citep{moe95}.
In this regard, several studies
of the spatial distribution of
novae in the nearby spiral M31 have suggested that the novae are 
primarily associated with the galaxy's bulge (e.g. Ciardullo et al. 1987;
Shafter \& Irby 2001; Darnley et al. 2006). These results were somewhat
surprising given the predictions of the Yungelson et al. (1997) population
synthesis models, but may have an alternative explanation. In particular,
Ciardullo et al. (1987) were first to put forward the idea that the high nova rate
in M31's bulge might be explained by a population of novae spawned in that
galaxy's globular cluster system and subsequently ejected into the bulge
by three-body interactions within the clusters, or by tidal disruption
of the clusters, or both. While extremely speculative, this process
might also explain the high nova rate found by Shara \& Zurek (2002) for M87,
a galaxy with an unusually high specific frequency of globular clusters.
Further study of the possible correlation of a galaxy's LSNR and
its specific frequency of globular clusters should be undertaken to
explore this intriguing possibility more thoroughly.

\section{Conclusions} 

A three-year H$\alpha$ survey of M94 spanning seven independent epochs
yields the detection of four nova candidates. One of the nova
candidates appeared to fade quite slowly, and may be another type of transient
source. Assuming all four nova candidates are in fact novae, we estimate the
annual nova rate of M94 to be 5.0$^{+1.8}_{-1.4}$ yr$^{-1}$, by using a Monte
Carlo simulation.  The corresponding
LSNR of M94 is 1.4 $\pm$ 0.5
novae per year per 10$^{10}L_{\odot,K}$, when the $K$ magnitude is obtained
from the $(B-K)$ color and the total $B$ magnitude.  When the $K$ magnitude
is derived from 2MASS the LSNR value is little changed at
1.5 $\pm$ 0.4 novae per year per 10$^{10}L_{\odot,K}$.

\indent Uncertainties in the measurement of nova rates in spiral galaxies
arise mainly due to our difficulty in accurately determining the effects of
extinction internal to the galaxy. The LSNR determined for M94 is slightly
below the
typical value of (2 -- 3) $\times$ 10$^{-10} L_{\odot,K}^{-1}$ yr$^{-1}$ for
galaxies with measured nova rates but is consistent within the uncertainty of
our measurements. Though the values for the Magellanic Clouds are about 2--3
times greater than the mean LSNR value, generally
there does not seem to be a strong
dependence of the LSNR with different Hubble types. The higher LSNRs of the
Magellanic Clouds provide support for the population synthesis models which
predict that low-mass, late-type galaxies are more likely to have a higher
production of novae than high-mass, early-type galaxies. 

\indent Future observations should focus on achieving more
frequent temporal sampling that will improve the determination of nova rates.
Dedicated surveys such as those carried out by PanStarrs and the LSST will
be very helpful in this regard. Such surveys will
not only help to obtain a larger sample of light curves to be used in the
Monte Carlo simulations but also reduce uncertainties in the measurements of
the effective survey time. It will also allow us to better assess the possible
effect of stellar population on nova light curve properties.

\vspace{0.25in}
\noindent We thank an anonymous referee for helpful suggestions to improve
the paper. A.W.S. gratefully acknowledges support through NFS
grant AST 06-07682.

\clearpage
\begin{deluxetable}{c c c c}
\tablewidth{0pt}
\tablecolumns{4}
\tablecaption{Summary of Observations}
\tablehead{
\colhead{} & \colhead{Julian Date} & \colhead{Number of} & \colhead{Total
  Integration Time} \\ \colhead{UT Date} & \colhead{(2,450,000+)} &
\colhead{Exposures} & \colhead{(hr)}}
\startdata
2005 May 01...................... & 3491.5 & 10 & 2.5 \\
2005 May 02...................... & 3492.5 & 2 & 0.5 \\
2005 May 30...................... & 3520.5 & 5 & 1.25 \\
2005 May 31...................... & 3521.5 & 3 & 0.75 \\
2006 Apr 19...................... & 3844.5 & 11 & 2.75 \\
2006 May 24...................... & 3879.5 & 11 & 2.0 \\
2006 Jun 17...................... & 3903.5 & 10 & 1.67 \\
2007 Mar 10...................... & 4169.5 & 18 & 3.0 \\
2007 Jun 08...................... & 4259.5 & 18 & 3.0 \\
\enddata
\end{deluxetable}

\clearpage
\begin{deluxetable}{cccccc}
\tablewidth{0pt}
\tablecolumns{6}
\tablecaption{Magnitudes and Positions of M94 Nova Candidates}
\tablehead{
\colhead{} & \colhead{Julian Date} & \colhead{$\alpha$} & \colhead{$\delta$} &
\colhead{$\Delta r$} & \colhead{$m_{\rm H \alpha}$}  \\ \colhead{Nova} & \colhead{(2,450,000+)} & \colhead{(J2000.0)} & \colhead{(J2000.0)} &
\colhead{(arcmin)} & \colhead{(mag)}}
\startdata
M94N~2005-05a & 3520.5 & 12 51 04.3 & 41 06 41 & 2.18 & 19.7 \\
M94N~2006-04a\tablenotemark{a} & 3844.5 & 12 50 54.4 & 41 09 08 & 1.94 & 20.7 \\ 
              & 3879.5 & \nodata & \nodata & \nodata & 21.3: \\
              & 3903.5 & \nodata & \nodata & \nodata & 21.2 \\
M94N~2006-04b & 3844.5 & 12 50 44.7 & 41 08 23 & 1.97 & 20.0 \\
              & 3879.5 & \nodata & \nodata & \nodata & 21.1: \\
              & 3903.5 & \nodata & \nodata & \nodata & 20.9 \\
M94N~2006-06a & 3903.5 & 12 50 36.8 & 41 04 20 & 4.22 & 19.8 \\
\enddata
\tablecomments{The units for right ascension are hours, minutes, and seconds.
  The units for declination are degrees, arcminutes, and arcseconds.  We
  assume these values to be accurate to $\sim 1\arcsec$.  The distance between
  the center of M94 and the nova is defined as $\Delta r$.}
\tablenotetext{a}{Slowly fading transient source that is possibly a long-period variable star.}
\end{deluxetable}

\clearpage

\clearpage

\clearpage

\clearpage


\begin{thebibliography}{}

\bibitem[Ciardullo et al.(1987)]{cia87} Ciardullo, R., Ford, H. C., Neill, J. D., Jacoby, G. H., \& Shafter, A. W.  1987, \apj, 318, 520
\bibitem[Alard \& Lupton (1998)]{ala98} Alard, C. \& Lupton, R. H. 1998, \apj, 503, 325
\bibitem[Ciardullo et al.(1990a)]{cia90a} Ciardullo, R., Ford, H. C.,
  Williams, R. E., Tamblyn, P., \& Jacoby, G. H.  1990a, \aj, 99, 1079
\bibitem[Ciardullo et al.(1990b)]{cia90b} Ciardullo, R., Shafter, A. W., Ford,
  H. C., Neill, J. D., Shara, M. M., \& Tomaney, A. B.  1990b, \apj, 356, 472
\bibitem[Coelho et al.(2008)]{coe08} Coelho, E. A., Shafter, A. W., \&
  Misselt, K. A.  2008, \apj, 686, 1261
\bibitem[Darnley et al. (2006)]{dar06} Darnley, M. J. et al. 2006, \mnras, 369, 257
\bibitem[Della Valle et al.(1994)]{del94} Della Valle, M., Rosino, L.,
  Bianchini, A., \& Livio, M.  1994, \aap, 287, 403
\bibitem[de Vaucouleurs et al.(1991)]{dev91} de Vaucouleurs, G., de
  Vaucouleurs, A., Corwin, H. G., Jr., Buta, R. J., Paturel, G., \& Fouque, P.
  1991, Third Reference Catalogue of Bright Galaxies (New York: Springer)
\bibitem[Ferrarese et al.(2003)]{fer03} Ferrarese, L., C\^{o}t\'{e}, P., \&
  Jord\'{a}n, A. 2003, \apj, 599, 1302
\bibitem[Gavazzi \& Scodegigo(1996)]{gav96} Gavazzi, G. \& Scodegigo, M. 1996 \aap, 312, L29
\bibitem[Herrmann et al.(2008)]{her08} Herrmann, K. A., Ciardullo, R.,
  Feldmeier, J. J., \& Vinciguerra, M.  2008, \apj, 638, 630
\bibitem[Jarrett et al.(2003)]{jar03} Jarrett, T. H., Chester, T., Cutri, R.,
  Schneider, S. E., \& Huchra, J. P.  2003, \aj, 125, 525
\bibitem[Johnson(1966)]{joh66} Johnson, H. L. 1966, \apj, 143, 187
\bibitem[Kolb(1995)]{kol95} Kolb, U. 1995, in Cataclysmic Variables,
  ed. A. Bianchini \& M. Della Valle (ASSL Vol. 205; Dordrecht: Kluwer), 511
\bibitem[Livio(1992)]{liv92} Livio, M.  1992, \apj, 393, 516
\bibitem[M\"ollenhoff et al.(1995)]{moe95} M\"ollenhoff, C., Matthias, M., \&
  Gerhard, O. E.  1995, \aap, 301, 359
\bibitem[Moses \& Shafter(1993)]{mos93} Moses, R. N., \& Shafter, A. W.  1993,
  \baas, 25, 1248
\bibitem[Neill \& Shara(2004)]{nei04} Neill, J. D., \& Shara, M. M.  2004,
  \aj, 127, 816
\bibitem[Oke(1974)]{oke74} Oke, J. B.  1974, \apjs, 27, 21
\bibitem[Prialnik et al.(1982)]{pri82}  Prialnik, D., Livio, M., Shaviv, G.,
  \& Kovetz, A.  1982, \apj, 257, 312
\bibitem[Ritter et al.(1991)]{rit91} Ritter, H., Politano, M., Livio, M., \&
  Webbink, R. F.  1991, \apj, 376, 177
\bibitem[Schegel et al.(1998)]{sch98} Schegel, D. J., Finkbeiner, D. P., \&
  Davis, M.  1998, \apj, 500, 525
\bibitem[Shafter et al.(2000)]{sha00} Shafter, A. W., Ciardullo, R., \&
  Pritchet, C. J. 2000, \apj, 530, 193
\bibitem[Shafter \& Irby(2001)]{sha01} Shafter, A. W., \& Irby, B. K.  2001,
  \apj, 563, 749
\bibitem[Shara et al.(1980)]{sha80} Shara, M. M., Prialnik, D., \& Shaviv, G.
  1980, \apj, 239, 586
\bibitem[Shara \& Zurek(2002)]{sha02} Shara, M. M., \& Zurek, D. R.  2002, in
  AIP Conf. Proc. 637, Classical Nova Explosions, ed. M. Hernanz \& J. Jose
  (Melville: AIP), 457
\bibitem[Tutukov \& Yungelson(1995)]{tut95} Tutukov, A. V., \& Yungelson,
  L. R.  1995, in Cataclysmic Variables, ed. A. Bianchini \& M. Della Valle
  (ASSL Vol. 205; Dordrecht: Kluwer), 495
\bibitem[Waller et al.(2001)]{wal01} Waller, W. H., Fanelli, M. N., Keel,
  W. C., Bohlin, R., Collins, N. R., Madore, B. F., Marcum, P. M., Neff,
  S. G., O'Connell, R. W., Offenberg, J. D., Roberts, M. S., Smith, A. M., \&
  Stecher, T. P.  2001, \aj, 121, 1395
\bibitem[Warner(1995)]{war95} Warner, B.  1995, Cataclysmic Variable Stars
  (Cambridge: Cambridge Univ. Press)
\bibitem[Williams et al.(2004)]{will04} Williams, G. G., Olszewski, E.,
  Lesser, M. P., \& Burge, J. H.  2004, Proc. SPIE, 5492, 787
\bibitem[Williams \& Shafter(2004)]{wil04} Williams, S. J., \& Shafter,
  A. W. 2004, \apj, 612, 867
\bibitem[Yungelson et al.(1997)]{yun97} Yungelson, L., Livio, M., \& Tutukov,
  A.  1997, \apj, 481, 127
\bibitem[Zwicky(1942)]{zwi42} Zwicky, F.  1942, \apj, 96, 28 

\end{thebibliography}
\end{document}